\documentclass[prl,footinbib,twocolumn,floatfix]{revtex4}

\usepackage{amsfonts,amsmath,amssymb}

\usepackage{tikz}

\newcommand{\ra}{\rightarrow}

\newcommand{\CC}{{\mathbb C}}
\newcommand{\ZZ}{{\mathbb Z}}

\newcommand{\PP}{{\mathbb P}}

\newcommand{\hG}{{\hat G}}

\newcommand{\Tr}{{\rm Tr}}
\newcommand{\cH}{{\mathcal H}}

\begin{document}

\title{Anomalies of discrete symmetries in three dimensions and group cohomology}

\author{Anton Kapustin}
\affiliation{California Institute of Technology, Pasadena, CA}
\author{Ryan Thorngren}
\affiliation{University of California, Berkeley, CA}

\begin{abstract}

We study 't Hooft anomalies for a global discrete internal symmetry $G$. We construct examples of bosonic field theories in three dimensions with a nonvanishing 't Hooft anomaly for a discrete global symmetry. We also construct field theories in three dimensions with a global discrete internal symmetry $G_1\times G_2$ such that gauging $G_1$ necessarily breaks $G_2$ and vice versa. This is analogous to the Adler-Bell-Jackiw axial anomaly in four dimensions and parity anomaly in three dimensions.

\end{abstract}

\maketitle

\section{Introduction}

Since the  discovery by Adler \cite{Adler} and Bell and Jackiw \cite{BJ} of the anomalous nonconservation of the axial current, anomalies have played an increasingly important role in particle physics. Recently anomalies found applications in condensed matter physics: they appear, implicitly or explicitly, in the classification of Symmetry Protected Topological (SPT) phases \cite{SPT,Wenanomalies}. This viewpoint sheds a new light on anomalies and leads to some surprising conclusions. Motivated by these developments, we study anomalies of global discrete internal symmetries.  In particular, we show that such anomalies can afflict bosonic field theories in odd space-time dimensions. 

There are several different but related kinds of anomalies. The original ABJ discovery \cite{Adler,BJ} was that a classical symmetry can be violated on the quantum level. We will call this phenomenon an ABJ anomaly. Anomalies can also affect gauge symmetries; gauge theories which suffer from such anomalies are inconsistent on the quantum level. Finally, it might happen that a global symmetry is consistent on the quantum level, but cannot be promoted to a gauge symmetry because the resulting gauge theory would be anomalous. In such a case one says that a global symmetry has an 't Hooft anomaly \cite{tHooft}. 't Hooft argued that 't Hooft anomalies of continuous symmetries are preserved under RG flow and thus constrain possible RG trajectories. 

The source of all these anomalies is chirality: either the theory itself is chiral, or the global symmetry acts in a chiral way. Since chiral matter exists only in even space-time dimensions, it is often said that in odd space-time dimensions anomalies are absent. However, in odd space-time dimensions there is another source of chirality, namely Chern-Simons couplings. This suggests that there may be anomalies whose existence depends on Chern-Simons interactions. It is this mechanism that causes the parity anomaly of  3d gauge theories with fermions  \cite{Redlich}.  In these theories, maintaining gauge-invariance  may require adding a Chern-Simons interaction which breaks parity. This is an ABJ anomaly for a global space-time symmetry. We will show that a similar mechanism can lead to anomalies for global discrete internal symmetries in bosonic theories.

\section{'t Hooft anomalies and group cohomology}

ABJ and 't Hooft anomalies are closely related, and it is instructive to address the latter first. 't Hooft anomaly for a global symmetry group $G$ is an obstruction for gauging $G$. If $G$ is a connected Lie group, the form of the 't Hooft anomaly is tightly constrained by the Wess-Zumino consistency conditions \cite{WZ}. They imply that in $d$ space-time dimensions possible anomalies for $G$ are classified by Chern-Simons actions in dimension 
$d+1$ \cite{Weinberg}. An intuitive reason for this is as follows. On a $d+1$-manifold with a boundary, the Chern-Simons actions is gauge-invariant only up to boundary terms. To cancel the boundary terms, one has to couple the bulk theory to a $d$-dimensional boundary theory with an 't Hooft anomaly for $G$.  This mechanism for canceling 't Hooft anomalies is called anomaly inflow. Conversely, if one assumes that an 't Hooft anomaly in $d$ dimensions can be canceled by an anomaly inflow from $d+1$ dimensions, then anomalies in $d$ dimensions must be classified by possible Chern-Simons actions in $d+1$ dimensions. 

If $G$ is not connected, Wess-Zumino consistency conditions are not as constraining. In the extreme case of  a finite symmetry group $G$, they become vacuous. As a substitute, let us assume that the anomaly can be canceled by inflow from $d+1$ dimensions. Then 't Hooft anomalies in $d$ dimensions should be classified by topological actions in $d+1$ dimensions \cite{Wenanomalies}. For general $G$, such actions are classified by elements of the abelian group $H^{d+2}(BG,\ZZ)$ \cite{DW}. Here $BG$ is the classifying space for $G$-bundles \footnote{Cohomology of $BG$ is also known as group cohomology of $G$. It should not be confused with the cohomology of $G$ regarded as a topological space.}. Typically, $BG$ is an infinite-dimensional space defined up to homotopy equivalence. For example, for $G=U(1)$ one can take $BG=\CC\PP^\infty$. For finite $G$ there is a nice explicit construction of $BG$ \cite{Milnor}. 

The classification of topological actions in terms of cohomology of $BG$ assumes that the action depends only on the gauge field. For fermionic systems the action may also depend on the spin structure, and then more complicated actions exist \cite{DW,Wenanomalies}. In this note we focus on bosonic systems and their anomalies. 

From now on $G$ will be a finite symmetry group. For finite $G$ we have an isomorphism $H^{d+2}(BG,\ZZ)\simeq H^{d+1}(BG,U(1))$ \cite{Evens}. Note that the same group classifies bosonic SPT phases with global symmetry $G$ in $d+1$ dimensions \cite{SPT}.  The reason for this is as follows. Group cohomology classification of SPT phases relies on gauging $G$ and integrating out everything except the gauge field for $G$. This gives an effective topological action for the $G$-connection, and such actions for finite $G$ are classified by $H^{d+1}(BG,U(1))$ \cite{DW}. Consequently, the boundary of an SPT phase classified by a cohomology class $\omega\in H^{d+1}(BG,U(1))$ must either break $G$ spontaneously or carry a field theory with a global symmetry $G$ which has an 't Hooft anomaly $\omega$ \cite{SPT,Wenanomalies}.  

The anomaly inflow assumption is very natural, but we do not know how to prove it rigorously. There might exist theories whose 't Hooft anomalies cannot be canceled by anomaly inflow and thus are not related to SPT phases. Since our main goal here is to produce new examples of anomalous theories, we leave this issue for future work.

\section{'t Hooft anomalies in three dimensions}

In $d=3$ the relevant cohomology group is $H^4(BG,U(1))$. This cohomology group is non-vanishing for  $G=\ZZ_n\times\ZZ_n$:  $H^4(B(\ZZ_n\times\ZZ_n),U(1))=\ZZ_n\times\ZZ_n$. (In contrast, $H^4(B\ZZ_n,U(1))=0$ for all $n$ \cite{Evens}.) Thus there should exist 3d field theories with a global symmetry group $\ZZ_n\times\ZZ_n$ which cannot be gauged.

To produce an example of such a theory we use the insight of Ref. \cite{VS} which argued that on a gapped boundary of an SPT phase the global symmetry $G$ must either be broken or realized projectively. That is, relations between generators of $G$ hold only modulo elements of a  gauge group $N$. This means that the symmetry of the theory is not a product $G\times N$, but an {\it extension} of $G$ by $N$. By definition, an extension of $G$ by $N$ is any group $\hG$ which has $N$ as a normal subgroup and such that $\hG/N=G$. From the physical viewpoint, $N$ has to be a normal subgroup, because conjugation by any element of $\hG$ must map a gauge symmetry to a gauge symmetry. We will be interested in the special case when $N$ is abelian, and $G$ acts trivially on the gauge fields. In this case every element of $\hG$ commutes with every element of $N$, and one says that $\hG$ is a central extension of $G$ by $N$. Central extensions are classified by the cohomology group $H^2(BG,N)$ \cite{Evens}.  Note that even if both $G$ and $N$ are abelian, $\hG$ may be nonabelian. 

Suppose that the full symmetry group $\hG$ is a central extension of $G$ by an abelian gauge group $N$. 
In general, the fact that $G$ is realized projectively on the fields  does not lead to 't Hooft anomalies. Let $N=\ZZ_2$, $\hG=\ZZ_4$ and $G=\ZZ_2$. Here $G$ is generated by a single element $g$ satisfying $g^2=n$, where $n$ is the generator of $N$. This is an example of a nontrivial extension of $\ZZ_2$ by $\ZZ_2$. It is easy to construct models where the global $\ZZ_2$ acts as above but nevertheless can be gauged. For example, consider a $U(1)$ gauge theory coupled to a pair of scalars which transform as a doublet of $U(2)$. The global symmetry of this model is $U(2)/U(1)=SU(2)/\ZZ_2$. Suppose further that the model has a singlet scalar of charge $2$ which condenses at a high energy scale and Higgses $U(1)$ to $\ZZ_2$. The global $SU(2)/\ZZ_2$ contains a $\ZZ_2$ subgroup generated by a transformation
\begin{equation}
g: \phi_\pm \mapsto \pm i \phi_\pm,\quad g^2=-1. 
\end{equation}
Clearly, there is no obstruction for gauging the whole $U(2)$, which includes the finite subgroup generated by $g$.  

On the other hand,  a projective action of $G$ opens a {\it possibility} for an 't Hooft anomaly. Indeed, suppose the gauge field for $N$ has a nontrivial topological action. As mentioned above, such actions are classified by elements of $H^3(BN,U(1))$. On the other hand, topological actions for $\hG$ are classified by elements of $H^3(B\hG,U(1))$. A 3-cocycle on $B\hG$ can always be restricted to the subspace $BN$ to give a 3-cocycle on $N$. However, in general not every 3-cocycle on $N$ can be obtained in this way. That is, the restriction map $r: H^3(B\hG,U(1))\ra H^3(BN,U(1))$ may fail to be onto. If the original theory has an action defined by $\omega\in H^3(BN,U(1))$ which is not in the image of $r$, such a theory will have an 't Hooft anomaly \footnote{We are grateful to E. Witten for suggesting this.}.

There is a simple example of such a situation. Let $N=\ZZ_3$ and $G=\ZZ_3\times\ZZ_3$. There exist both abelian and nonabelian extensions of $G$ by $N$, but it turns out that only nonabelian extensions have 't Hooft anomaly. There are two nonequivalent nonabelian extensions $\hG$. One of them is the finite Heisenberg group $\cH_{27}$ of order $27$ with generators $x,y,z$ and relations
\begin{equation}
x^3=y^3=z^3=1,\ yx=zxy,\ xz=zx,\ yz=zy.
\end{equation}
The other one also has three generators $x,y,z$ but different relations:
\begin{equation}
x^3=y^3=z,\ z^3=1, \ yx=zxy,\ xz=zx,\ yz=zy.
\end{equation}
Both of these groups have a faithful three-dimensional representation and therefore can be thought of as subgroups of $U(3)$. For example, to embed the Heisenberg group into $U(3)$, we can set
\begin{equation}\label{repthree}
x=\begin{pmatrix} 1 & 0 & 0\\  0 &\eta & 0\\ 0 & 0 & \eta^2\end{pmatrix},\ y=\begin{pmatrix} 0 & 1 & 0\\ 0 & 0& 1\\ 1 & 0 & 0\end{pmatrix},\ 
z=\begin{pmatrix} \eta & 0 & 0\\ 0 & \eta & 0\\ 0 & 0 & \eta\end{pmatrix},
\end{equation}
where  $\eta=\exp(2\pi i/3)$.
That is, $x$ and $y$ are clock and shift matrices, respectively, while $z$ is the generator of the ``scalar'' $\ZZ_3$ subgroup. Using the computer algebra package HAP \cite{HAP}, we checked that for both groups the restriction map $r$ is zero, so a nontrivial action for the $\ZZ_3$ gauge field cannot be promoted to a nontrivial action for a $\hG$ gauge field.  

One can prove a general theorem about extensions of $\ZZ_p\times\ZZ_p$ by $\ZZ_p$, where $p$ is an odd prime, which says that all nonabelian extensions are afflicted by an 't Hooft anomaly, while abelian ones are not \cite{KRtoappear}. Since the proof is rather elaborate, we prefer to give a heuristic argument which leads to the same conclusion and also shows that for $p=2$ the 't Hooft anomaly vanishes. 

Suppose the finite group $\hG$ embeds into a connected Lie group $H$ which is also a symmetry of the theory. Then instead of gauging $\hG$ we can try to gauge $H$, which is a more familiar task.  We must also make sure that when the action is restricted to the subgroup $N$, it reduces to the original action.  If this turns out impossible, this does not prove that $\hG$ cannot be gauged. But if the failure persists for all natural choices of $H$, this is suggestive of  an 't Hooft anomaly. On the other hand, if $H$ can be gauged, then so can $\hG$. 

Let us review topological actions for $G=\ZZ_n$. Such actions are classified by $H^3(\ZZ_n,U(1))=\ZZ_n$ and are constructed as follows \cite{BanksSeiberg}. We regard $\ZZ_n$ as a subgroup of $U(1)$. Accordingly, a topological action has the form
\begin{equation}
S=\frac{\ell}{2\pi}\int_X a da+\frac{n}{2\pi} \int_X b da,
\end{equation}
where $a$ and $b$ are $U(1)$ gauge fields and $\ell$ is an integer. The field $b$ is the dual of a charge-$n$ scalar whose expectation value breaks $U(1)$ down to $\ZZ_n$. A field transformation $b\mapsto b+a$ is equivalent to a shift $\ell\mapsto \ell+n$. Hence $\ell$ is defined modulo $n$, in agreement with the group cohomology classification. On the other hand, the usual $U(1)$ Chern-Simons action at level $k$ is 
\begin{equation}
\frac{k}{4\pi} \int_X a da.
\end{equation}
Hence upon Higgsing $U(1)$ down to $\ZZ_n$,  $U(1)$ Chern-Simons theory at level $k$ reduces to $\ZZ_n$ topological theory at level $\ell=k/2$. Note that $\ell$ is half-integral for odd $k$. This is because for odd $k$ the $U(1)$ Chern-Simons theory implicitly depends on spin structure, and so does the $\ZZ_n$ theory obtained by Higgsing \cite{DW}. To get a topological $\ZZ_n$ gauge theory the $U(1)$ Chern-Simons level must be even.

We are now ready to consider embeddings of $\hG$ into a Lie group $H$. For definiteness, consider the case of the finite Heisenberg group $\cH_{27}$. The natural choice is $H=U(3)$, with $x,y,z$ as in (\ref{repthree}). The gauge subgroup $N$ is generated by $z$. The most general Chern-Simons actions for $U(3)$ is given by
\begin{equation}
\frac{k}{4\pi}\int_X \Tr \left(A dA+\frac{2}{3} A^3\right),
\end{equation}
where $k$ is integral and $\Tr$ is the trace in the fundamental representation. In order to get a topological action rather than a spin-topological one, $k$ has to be even \cite{DW}. When this action is restricted to the subgroup $U(1)$ consisting of scalar matrices, this action becomes $U(1)$ Chern-Simons action at level $3k$. Hence the $\ZZ_3$ subgroup generated by $z$ has level $\ell=3k/2=0\ {\rm mod}\ 3$. Therefore a nontrivial topological action for a $\ZZ_3$ gauge field cannot be promoted to a Chern-Simons action for a $U(3)$ gauge field. This suggests that $G$ has an 't Hooft anomaly.

On the other hand, there is no anomaly in the superficially very similar case $N=\ZZ_2$ and $G=\ZZ_2\times\ZZ_2$. In this case there are also two inequivalent nonabelian extensions of $G$ by $N$: the finite Heisenberg group $\cH_8$ with generators $x,y,z$ satisfying
\begin{equation}
x^2=y^2=z^2=1,\ xy=zyx,\quad xz=zx,\ yz=zy,
\end{equation}
and the group $Q_8$ with generators $x,y,z$ satisfying
\begin{equation}
x^2=y^2=z,\ z^2=1, \ xy=zyx,\ xz=zx,\ yz=zy.
\end{equation}
$\cH_8$ is isomorphic to the dihedral group of order $8$. For definiteness, consider the group $\cH_8$. It can be embedded into $U(2)$, so that $x$ and $y$ are represented by $2 \times 2$ clock and shift matrices, while $z=-1$.  Chern-Simons action  for $U(2)$ at level $k$ restricts to Chern-Simons action for $U(1)$ at level $2k$. Therefore the topological action for the $\ZZ_2$ subgroup generated by $z$ will have level $\ell=2k/2=k$. To get a nontrivial action for the $\ZZ_2$ subgroup one can take $k=1$. Thus even when the action for $N=\ZZ_2$ is nontrivial, one can promote $\ZZ_2$ to $U(2)$ at level $1$. Higgsing $U(2)$ down to $\cH_8$, one gets a consistent action for the $\cH_8$ gauge field which extends the original action for the $\ZZ_2$ gauge field. Note that for odd $k$ the $U(2)$ Chern-Simons theory is spin-topological rather than topological, so it is not clear from this argument whether the resulting $\cH_8$ gauge theory depends on spin structure or not. Using the package HAP, we checked that the restriction map $H^3(B\cH_8,U(1))\ra H^3(B\ZZ_2,U(1))$ is onto, and hence $\ZZ_2$ can be promoted to $\cH_8$ without introducing dependence on spin structure. The same is true for the other nonabelian extension $Q_8$. 

We can now produce a concrete example of a 3d QFT with an 't Hooft anomaly for an internal global discrete symmetry. We take three scalar fields and couple them to a $U(1)$ gauge field at level $\pm 2$. All three matter fields have charge $1$. If we wish, we can break $U(1)$ down to a $\ZZ_3$ subgroup by condensing yet another field of charge $3$. Let us postulate that all other interactions of the scalar fields have $G=\ZZ_3\times\ZZ_3$ symmetry whose generators acts on the three scalars as clock and shift matrices (\ref{repthree}). The above discussion implies that $G$ cannot be gauged, i.e. it has an 't Hooft anomaly.

\section{ABJ anomalies in three dimensions}

ABJ anomalies are closely related to 't Hooft anomalies. For example, one can interpret the famous axial anomaly as arising from an 't Hooft anomaly for $U(1)_{em}\times U(1)_A$. In general, suppose a global symmetry group has the product form $G=G_1\times G_2$. Suppose also that $G$ has a nontrivial 't Hooft anomaly, while both $G_1$ and $G_2$ have a trivial 't Hooft anomaly. Mathematically, this means that the cohomology class $\omega\in H^4(BG,\ZZ)$ becomes trivial when restricted to $BG_1$ or $BG_2$ (for example, because $H^4(BG_1,\ZZ)=H^4(BG_2,\ZZ)=0$). Then $G_1$ or $G_2$ can be gauged, but not the whole $G$. What is the fate of $G_2$ in a theory with a gauged $G_1$? Either it is still a global symmetry or it is not. If it is a symmetry, then it must have a nontrivial 't Hooft anomaly. But if $H^4(BG_2,\ZZ)$ is trivial, this is impossible. Hence $G_2$ is not a symmetry, i.e. gauging $G_1$ necessarily breaks $G_2$,  regardless of the precise form of the action for the $G_1$ gauge field. Conversely, gauging $G_2$ necessarily breaks $G_1$. We may regard this as a form of ABJ anomaly. 

The above example with $G=\ZZ_3\times\ZZ_3$ has exactly the right structure, since $H^4(B\ZZ_3,\ZZ)=0$. We can see how the ABJ anomaly arises by embedding $\hG$ into $U(3)$ as in (\ref{repthree}). Then $N=\ZZ_3$ is generated by the scalar matrix $z$, while $G_1$ is generated by the diagonal ``clock'' matrix $x$. We can regard $N$ and $G_1$ as commuting $\ZZ_3$ subgroups of $U(1)^3\subset U(3)$. To get a gauged action for $N\times G_1$, we need to pick Chern-Simons levels for the three $U(1)$ factors and then Higgs $U(1)^3$ down to $\ZZ_3\times\ZZ_3$ by condensing some scalars. If the level $\ell$ for the subgroup $N$ is $\pm 1\ {\rm mod}\ 3$, then the $U(1)^3$ levels must satisfy $k_1+k_2+k_3=\pm 2 \ {\rm mod}\ 3$. Such a triplet of integers will necessarily break the symmetry $G_2=\ZZ_3$ which cyclically permutes the three $U(1)$ factors. Since the breaking of $G_2$ is due to topological terms in the action, it can be regarded as a quantum effect analogous to the ABJ anomaly.  The analogy with the parity anomaly of 3d gauge theories is even closer. There gauge-invariance forces one to choose the Chern-Simons level $k$ to be half-integral, excluding the parity-invariant value $k=0$ \cite{Redlich}. Similarly, in our example $N\times G_1$ gauge-invariance together with the requirement that the topological action for the subgroup $N$ be nontrivial forces us to break the remaining global $\ZZ_3=G_2$.

\section{Concluding remarks}

't Hooft anomalies for continuous symmetries are not affected by the RG flow and therefore lead to constraints on the IR behavior of theories \cite{tHooft}. This is true even if the symmetry in question is spontaneously broken, since breaking a continuous symmetry results in Goldstone bosons, and an 't Hooft anomaly gives rise to Wess-Zumino-Witten terms in their effective action \cite{Weinberg}. 

For discrete symmetries the situation is different, since spontaneous breaking of discrete symmetries does not lead to massless particles. A phase with a discrete global symmetry $G$ broken down to nothing is always possible and has trivial 't Hooft anomaly, regardless of the 't Hooft anomaly of the UV theory. But if some subgroup $G_0\subset G$ remains unbroken, the same argument as for continuous symmetries shows that 't Hooft anomalies for $G_0$ must be the same in the UV and the IR. 

Further, we saw that even massive QFTs can have nontrivial 't Hooft anomalies, if at long distances they reduce to sufficiently complicated TQFTs. Thus a nonvanishing 't Hooft anomaly in general does not rule out a gapped phase with an unbroken symmetry: it only rules out a trivial gapped phase with an unbroken symmetry. 

Note that in \cite{VS} it was proposed that phases with this property (gapped phases with a symmetry $G$ which, while unbroken, requires topological order) are precisely the gapped surface phases of bosonic SPTs in one dimension higher. But we saw above that such phases may have trivial 't Hooft anomalies.  One such example is $G=\ZZ_2\times\ZZ_2$, $N=\ZZ_2$ and $\hG=\cH_8$ or $Q_8$. In this case, $G$ acts projectively, hence $N$ cannot be Higgsed without breaking some of $G$. Topological order in such theories is ``protected'' by the global discrete symmetry $G$. Nevertheless, such theories have trivial 't Hooft anomalies and therefore the corresponding SPT phases in four dimensions are trivial in the group cohomology classification. 

We are grateful to T. Senthil, E. Witten, N. Seiberg, V. Ostrik and P. Etingof for discussions. The work of A. K. was supported in part by the DOE grant  DE-FG02-92ER40701.


\begin{thebibliography}{99}

\bibitem{Adler} S.~L.~Adler,  Phys.\ Rev.\  {\bf 177}, 2426 (1969).

\bibitem{BJ} J.~S.~Bell and R.~Jackiw,  Nuovo Cim.\ A {\bf 60}, 47 (1969).

\bibitem{SPT} X.~Chen, Z.~-C.~Gu, Z.~-X.~Liu and X.~-G.~Wen,
  Phys.\ Rev.\ B {\bf 87}, 155114 (2013);  L.~-Y.~Hung and X.~-G.~Wen,
 Phys.\ Rev.\ B {\bf 87}, no. 16, 165107 (2013).
 
\bibitem{Wenanomalies} X-G.~Wen, Phys.\ Rev.\ D {\bf 88}, 045013 (2013). 


\bibitem{tHooft} G.~'t Hooft, 
  NATO Adv.\ Study Inst.\ Ser.\ B Phys.\  {\bf 59}, 135 (1980).

\bibitem{Redlich} A.~N.~Redlich, 
 Phys.\ Rev.\ Lett.\  {\bf 52}, 18 (1984).
 
\bibitem{WZ} J.~Wess and B.~Zumino,
  Phys.\ Lett.\ B {\bf 37}, 95 (1971).
  
\bibitem{Weinberg} S.~Weinberg, ``The quantum theory of fields. Vol. 2: Modern applications,''
 Cambridge, UK: Univ. Pr. (1996).


\bibitem{DW} R.~Dijkgraaf and E.~Witten,
  Commun.\ Math.\ Phys.\  {\bf 129}, 393 (1990).
  
\bibitem{Milnor} J.~Milnor, Ann.\ Math.\ (2) {\bf 63}, 430 (1956).
  
\bibitem{Evens} L.~Evens, ``The Cohomology of Groups,'' Oxford University Press, 1991.

\bibitem{VS} A.~Vishwanath and T.~Senthil,
  Phys.\ Rev.\ X {\bf 3}, 011016 (2013)
  
\bibitem{HAP} HAP - Homological Algebra Programming package, available at http://hamilton.nuigalway.ie/Hap/www/

\bibitem{KRtoappear} A. Kapustin and R. Thorngren, in preparation.

\bibitem{BanksSeiberg} T.~Banks and N.~Seiberg,
  Phys.\ Rev.\ D {\bf 83}, 084019 (2011)


  
\end{thebibliography}
\end{document}